# Optimizing thermoelectric performances of low-temperature SnSe compounds by electronic structure design

A. J. Hong[a], L. Li[a], H. X. Zhu[a], Z. B. Yan[a], J. –M. Liu*[a], and Z. F. Ren[b]

**[Abstract]** Recently SnSe compound was reported to have a peak thermoelectric figure-of-merit (*ZT*) of 2.62 at 923 K, but the *ZT* values at temperatures below 750 K are relatively low. In this work, the electronic structures of SnSe are calculated using the density functional theory, and the electro- and thermo-transport properties upon varying chemical potential (or carrier density) are evaluated by the semi-classic Boltzmann transport theory, revealing that the calculated *ZT* values along the a- and c-axes below 675 K are in agreement with reported values, but that along the b-axis can be as high as 2.57 by optimizing the carrier concentration to ~$3.6\times10^{19}$ cm$^{-3}$. It is suggested that a mixed ionic-covalent bonding and heavy-light band overlapping near the valence band are the reasons for the higher thermoelectric performance.

## I. INTRODUCTION

Thermoelectric (TE) materials and devices have been receiving much attention over past few decades due to their capabilities to commit direct and reversible conversion between electrical energy and heat energy.[1-10] Searching proper TE materials with high conversion efficiency is important for developing advanced TE technologies.[11] For a TE device, the thermal power $PF=S^2\sigma$ and the dimensionless figure of merit $ZT=S^2\sigma T/k_{tot}$ are two core parameters,[1] where $S$, $T$, $\sigma$, $k_{tot}$, $k_l$, and $k_e$ represent respectively the Seebeck coefficient, absolute temperature, electrical conductivity, total thermal conductivity, and lattice and electronic components of $k_{tot}$ ($k_{tot}=k_l+k_e$). Nevertheless, for realistic TE materials, achieving high *PF* and low $\kappa_{tot}$ simultaneously remains to be a contradictory issue due to the competing dependences of the parameters ($S$, $\sigma$, $\kappa_l$, and $\kappa_e$) on chemical and electronic structures,[5,9,10] making a maximization of the TE performances extremely challenging. Just because of such complexities, theoretical predictions from the full-scale first-principles calculations have been of interest for guiding TE materials synthesis.

It has been proposed that a combination of electronic crystal and phononic glass in one material is a major approach.[1, 3-7] This requires delicate design of the crystalline and electronic structures simultaneously. The electrical conductivity $\sigma$ depends on the carrier density and mobility which are both determined by electronic structure, while the Seebeck coefficient $S$ is essentially determined by the gradient of density of states (DOS) of the conduction band near the Fermi level (chemical potential).[11] Surely, the lattice thermal conductivity $\kappa_l$ can be substantially suppressed by modulating the material microstructure,[3] but the electronic thermal conductivity $\kappa_e$ is again highly dependent of the electronic structure. In a general sense, one is in a good position to optimize the TE performance of a material by tentatively designing the electro- and thermo-transport behaviors based on the electronic structure. In the other words, the electronic structure engineering has become a major branch of TE materials science.

To engineer the transport behaviors, one of the effective and often employed approaches is to modulate the carrier density ($n$) by chemical substitution and charge doping (thus varying chemical potential $m$) without seriously distorting the topology of electronic structure. Such a carrier density modulation may lift or lower the Fermi level (i.e. $m$) so that the ZT can be optimized. A common strategy is to choose some TE parent compounds with one or more outstanding TE parameters and then to improve the other parameters in order to reach an optimization of the PF and ZT values. Along this line, a series of parent compounds have been explored, followed by extensive investigations on the consequence of various doping trials for each compound.

Recent studies[9,12-14] revealed that selenium-based compounds are promising TE candidates. While polycrystalline $b$-CuAgSe exhibited very low $k_{tot}$ (<0.5Wm$^{-1}$K$^{-1}$) and the ZT reaches up to ~0.95 at 623 K[14], polycrystalline Cu$_2$Se was found to have a ZT value as high as 1.5 at $T$=1000K.[12] Surprisingly, single crystal SnSe compound was reported to have its ZT value as high as 2.62 along the $b$-axis and 2.30 along the $c$-axis at $T$~923K.[9] These values represent the highest ones reported so far, and thus allow SnSe to be a good platform for exploring the possibilities for even better TE performances. Similar to Cu$_2$Se, the SnSe compound has two phases.[15,16] The high-$T$ phase possesses ultralow $\kappa_l$ (<0.25Wm$^{-1}$K$^{-1}$ at $T$>800K) and moderate PF values, thus leading to surprisingly high ZT values.[9] Very differently, the low-$T$ phase, however, shows much lower TE properties and its ZT value along the $b$-axis is only ~1.0 at $T$~750K, the transition point between the two phases.

Considering the requirement for intermediate-$T$ applications and the fact that the low-$T$ SnSe compound has high Seebeck coefficient ($S$>500μVK$^{-1}$ at $T$<675K and $S$>400μVK$^{-1}$ at $T$=300~750K), a promotion of its overall TE performances is highly appealed. Experimentally reported carrier density of SnSe alloys in the intermediate *T*-range is relatively low ($n \sim 10^{18} cm^{-3}$), [9] which leaves sufficient space for carrier density modulation by means of low-level carrier doping.

## II. APPROACH AND THEORY

In this work, we intend to perform a full-scale first-principles electronic structure calculation on the low-*T* SnSe phase, and subsequently evaluate the electro- and thermo-transport properties at finite *T* by means of the semi-classic Boltzmann transport theory. In this scheme, all the thermodynamic properties relevant to electron transport can be calculated quantitatively so long as chemical potential *m* (or doping carrier density *n*) is given. Along this line, one is allowed to investigate the consequence of charge/carrier doping (low level) without seriously distorting the topology of electronic structure. Similar schemes have been extensively employed for TE materials design and performances optimization.[17-20] Our calculations indeed predict the significant impact of carrier density variation on the TE properties of low-*T* SnSe phase.

We start from experimentally determined crystal structure of the low-*T* SnSe phase.[15] It has the space group Pnma (#62), as shown in Fig.1, and the lattice constants are *a*=11.58Å, *b*=4.22Å, *c*=4.40Å. The Sn and Se atoms are located on two different planes with the dihedral angle of ~20°. We employ the density functional theory (DFT) scheme with full-potential linearized augmented plane-wave (LAPW) method implemented.[21] The WIEN2K package can offer high-precision and accurate calculations on electronic structure with relatively low efficiency. The exchange and correlation interactions are described using the generalized gradient approximation (GGA) and the Perdew-Burke-Ernzerhof (PBE) functional modified by Becke-Johnson potential (mBJ).[22] The muffin-tin radii are set as 2.5 a.u. for both Sn and Se atoms with well-converged basis set determined by RMT×Kn=7.0, corresponding to 5907 plane waves. The lattice structure is optimized by minimization of the forces (0.5 mRy/a.u.) acting on the atoms with a fixed lattice constant.

Given the whole set of electronic structure data and varying *m*, we employ the semi-classic Boltzmann transport theory to calculate the electro- and thermo-transport behaviors. The calculated transport coefficients are found to be converged using a shifted 20×55×53k mesh. The original *k*-mesh is interpolated onto a mesh five times as *k* dense. All of the calculations are implemented by solving the Boltzmann transport equation with the constant relaxation time approximation.[23] In details, the electrical conductivity tensors and electronic thermal conductivity tensors at non-zero electric current are obtained by the following equations: [23]

$$\sigma_{\alpha\beta}(T,\mu) = \frac{1}{\Omega}\int \sigma_{\alpha\beta}(\varepsilon)\left[-\frac{\partial f}{\partial \varepsilon}\right]d\varepsilon \qquad (1)$$

$$\kappa^0_{\alpha\beta}(T,\mu) = \frac{1}{e^2 T\Omega}\int \sigma_{\alpha\beta}(\varepsilon)(\varepsilon-\mu)^2\left[-\frac{\partial f}{\partial \varepsilon}\right]d\varepsilon \qquad (2)$$

where e, $\Omega$, $e$, and $f$ are the electron charge (e), reciprocal space volume ($\Omega$), carrier energy ($e$), and Fermi distribution function ($f$), respectively. The conductivity tensors $\sigma_{\alpha\beta}$ can be expressed as:

$$\sigma_{\alpha\beta}(\varepsilon) = \frac{e^2\tau}{N}\sum_{i,k} v_\alpha v_\beta \frac{\delta(\varepsilon-\varepsilon_{i,k})}{d\varepsilon} \qquad (3)$$

here, $t$ and $v_a$ ($v_b$) are the relaxation time and electron group velocity, $k$ is the wave-vector. In the standard procedure, the $S$ and $k_e$ at zero electric current can be obtained:

$$S_{ij}(T,\mu) = (\sigma^{-1})_{\alpha i} c_{\alpha j} \qquad (4)$$

$$\kappa_{ij} = \kappa^0_{ij} - T c_{i\beta}(\sigma^{-1})_{\alpha\beta} c_{\alpha j} \qquad (5)$$

where

$$c_{\alpha\beta}(T,\mu) = \frac{1}{eT\Omega}\int \sigma_{\alpha\beta}(\varepsilon)(\varepsilon-\mu)\left[-\frac{\partial f}{\partial \varepsilon}\right]d\varepsilon. \qquad (6)$$

It should be mentioned that relaxation time $t$ is weakly dependent of the band index and $k$-direction, and thus spatially anisotropic for most cases. However, this dependence is quite trivial and can be neglected safely without damaging much the results even in quantitative sense. Earlier studies[24,25] did indicate that the $t$ is orientation-independent for most materials, i.e. approximately isotropic. Even for superconducting cuprates whose electrical conduction is known to be anisotropic, this relaxation time remains almost isotropic. Therefore, realistic calculations often take this approximation, although these properties may be spatially anisotropic.

As well known, doping carrier density $n$ is given by:

$$n = n_0 - \int f(\varepsilon, \mu, T) D(\varepsilon) d\varepsilon \qquad (7)$$

where $n_0$ is valence electron number and $D(e)$ is the total DOS as a function of $e$ as evaluated from the electronic structure. There is a one-to-one correspondence between doping carrier density $n$ and chemical potential $m$ at a given $T$. To this stage, a self-consistent calculation based on the Boltzmann transport theory is immediate and no details of the practical calculation procedure will be repeated here.

## III. RESULTS AND DISCUSSION

### A. Electronic structures

The calculated band structures and DOS along the high symmetry lines are shown in Fig.2. The reciprocal space $a^*$, $b^*$, $c^*$ axes are parallel to the $a$, $b$, $c$ real space directions. All energies are referenced to the middle of the band gap. It can be seen that the conduction band minimum (CBM) is located at the $\Gamma$ point and there is a local CBM at (0.000, 0.329, 0.000) along the $\Gamma$-$Y$ line. The first and second valence band maximums (VBMs) are located at (0.000, 0.000, 0.354) and (0.000, 0.000, 0.444) along the $\Gamma$-$Z$ line. There is the third VBM at (0.000, 0.316, 0.000) along the $\Gamma$-$Y$ line, implying two indirect band gaps $E_g$=0.849 and 0.862eV along the $\Gamma$-$Y$ line. Along the $\Gamma$-$Z$ line, there are also two indirect band gaps $E_g$=0.842 and 0.828eV. The similar gaps allow similar transport properties along the $a$- and $b$- axes. As well known, for an indirect bandgap semiconductor, electrons cannot shift from the VBM to the CMB without momentum change. Fig.2 shows that the VBM and CBM have nearly equal momentum at the gap，. Namely, electrons are more easily excited into the conduction band at the band gap，than that at other band gaps.

It is noted that the gap，($E_g$=0.862) almost equals to previously measured value of 0.86eV.[9] We then focus on the upper part of the highest valence band. The dispersion along the $\Gamma$–$Z$ line is greater than that along the planar $A$–$Z$. The flat part of the highest valence band near the VBM is beneficial for high Seebeck and the conductivity is determined mostly by the steep band. In fact, earlier work[26-29] indicated that a mixture of the heavy and light bands near the valence edge is favorable for high TE performance, because the light band allows good electrical conduction and the heavy band benefits to high Seebeck coefficient. In addition, Fig.2 shows that the total DOS increases more rapidly near the VBM than that near the CBM. The major DOS contribution to the CBM comes from the Sn atoms while the Se atoms contribute more to the DOS near the VBM.

The orbital-decomposed band structures are presented in Fig.3(a) and (b) where the coarseness of curves scales the DOS intensity. It is seen that the VBM mainly comes from the Se $4p$ states, and the Sn $5p$ states only have weak contribution. This suggests that the $p$-type doping at the Sn site will increase the carrier density so as to improve the electrical conductivity, while the VBM shape can be roughly maintained so as to keep the high Seebeck coefficient. The projected DOS in the [-10.0eV, 10.0eV] interval is shown in Fig.3(c). A comparison of Fig.3(c) with Fig.2 allows us to conclude that the bands from the Fermi level to 5.0eV is mainly from the Sn $5p$ and Se $4p$ states, while the bands from -5.0eV to the Fermi level primarily comes from the Se $4p$ states. The highest bonding peaks near the VBM show the characteristics of the Se $4p$ electrons but also contain contributions of the Sn $5s$ electrons, implying the week $s$-$p$ hybridization between the Sn and Se atoms. Such $s$-$p$ hybridization can lead to dramatic DOS variation near the VBM, favorable for a high Seebeck coefficient.

We also calculate the electron density and charge density difference on the Se-Se-Se plane (schematically shown in Fig.1 by the shadow plane), as displayed in Fig.4(a) and (b) respectively. The Sn-Se bond has weak covalent component, which again confirms the $s$-$p$ hybridization between the Se $4p$ and Sn $5s$. The weak tendency for Se atoms to accumulate charge from the surrounding Sn atoms can be seen too in Fig.4(b). It is thus suggested that the bond between the Sn and Se atoms is more or less of mixed ionic-covalent nature.

**B. Thermoelectric properties**

Subsequently, we investigate the electro- and thermo-transport behaviors. In Fig.5(a)~(e) are plotted several calculated parameters as a function of $n$ respectively along the three major axes at $T$=675K. The $S(n)$ curves show the single-peaked pattern and the peak location and height shift along different major axes. The peak values of $S(n)$ along the $a$-, $b$- and $c$-axes reach up to 544.07, 690.37, and 655.13μV/K respectively, at $n$~$9.8\times10^{19}$, $2.3\times10^{19}$, $2.3\times10^{19}$cm$^{-3}$. For calculating other parameters, relaxation time $t$ is needed, but extracting its value from the *ab-initio* calculation is still challenging. Usually, the constant relaxation time approximation is used,[30-31] and we take $t$=0.5×10$^{-14}$s in the present calculations. The $S(n)$ dependences along the three axes are all monotonous at high doping level with small difference along the b- and c-axes, but the $s(n)$ along the a-axis is decreasing with increasing n at low doping level. It can be seen from Fig.5(a) that the $s(n)$ along the b-axis is much larger than along the a-axis at high doping level. It is noted that the $S(n)$ and $s(n)$ exhibit the opposite dependences, resulting in the *PF* peaks along these axes roughly at $n \sim 10^{20} \sim 10^{21} cm^{-3}$. The *PF* along the b-axis is about twice as large as that along the c-axis and is almost ten times larger than that along the a-axis. The $k_e(n)$ along all three axes is very sensitive to n. Fig.5(d) presents that the $k_e$ along the b- and c-axes first decreases and then increases with increasing n. However, the $k_e$ along the a-axis first increases and then decreases, and again rapidly increases with increasing n. In essence, the $k_e(n)$ dependence is determined by the band structure of SnSe crystals.

As an example, the evaluated $ZT(n)$ curves along the three major axes at $T=675K$ for the p-type doped systems are presented in Fig.5(e), where measured $k_l$ value was took from Ref.[9]. One sees that the *ZT* is sensitive to the p-type carrier density and a variation of n over two orders of magnitude is sufficient to modulate the *ZT* between the minimal and maximal. The $ZT(n)$ dependence is also anisotropic, yielding the relation $ZT_{b\text{-axis}} > ZT_{c\text{-axis}} > ZT_{a\text{-axis}}$ in agreement with experimental results.[9] The predicted highest *ZT* value of ~2.57 occurs at the p-type carrier density $n \sim 3.6 \times 10^{19} cm^{-3}$ along the b-axis, which appeals for experimental checking. We also predict the $ZT(n)$ of polycrystalline SnSe at the same temperature, as shown in Fig.5(e). The highest *ZT* value can reach up to 1.86 at the p-type carrier density $n \sim 4.2 \times 10^{19} cm^{-3}$.

### C. Comparison with experiments

Finally, we compare our calculated data with measured data. So far, measured *S*, *s*, and, $k_{tot}$ data along the three major axes of SnSe single crystals as a function of *T* are available.[9] For such a comparison, one needs measured $n(T)$ or $m(T)$ data for our calculations. Given the data in Ref.[9] and the assumption of constant n ($=5 \times 10^{17} cm^{-3}$) over the whole T-range, the as-evaluated $S(T)$ and $s(T)$ data are plotted in Fig.6(a)~(c) and Fig.6(d)~(f), focusing on the T-range from 300K to 700K. The calculated data coincide reasonably well with measured data along all the three axes, although the discrepancy becomes remarkable at both extremes of the

$T$-range, particularly for $s(T)$ along the $b$- and $c$-axes. The discrepancy is believed to most likely originate from the assumption of a constant relaxation time. In addition, we extract the measured $k_{tot}(T)$ data in Ref.[9] for evaluating the $ZT(T)$ data, and the results are presented in Fig.6(g)~(i) in comparison with measured $ZT(T)$ data.[9] Again, we see consistency between the calculated and measured $ZT$ data particularly in the low-$T$ range.

The capability of the present computational scheme may be highly appreciated considering the current status of quantitative predictions for TE performances. This allows comprehensive design and evaluation of the TE performances for a realistic material. One can always start from stoichiometric compound for electronic structure calculation, and then optimize the TE parameters by carefully tuning the chemical potential on condition of relative low-level doping so that the electronic structure topology remains qualitatively unchanged. This strategy is no doubt helpful for guiding practical synthesis and substitution/doping processes for better TE materials and performances.

## IV. SUMMARY

In summary, we have calculated the electronic structure and TE properties of low-$T$ SnSe compound using the first-principles calculations plus the semi-classic Boltzmann transport theory. It is revealed that the high Seebeck coefficient and good electrical conductivity are attributed to the $s$-$p$ hybridization and the mixed heavy-light band structure near the VBM. It is predicted that a proper modulation of the chemical potential or $p$-type carrier density can remarkably enhance the power factor $PF$ and figure-of-merit factor $ZT$. The calculated results are well consistent with experimental data reported recently. When the $p$-type carrier density is enhanced to ~$3.6 \times 10^{19}$cm$^{-3}$, the optimal $ZT$ values up to ~2.57 along the $b$-axis at $T$=675K are predicted.

**Acknowledgement:**

This work is supported by the National 973 Projects of China (Grant Nos. 2015CB654602), the Natural Science Foundation of China (Grant No. 51431006), and "Solid State Solar Thermal Energy Conversion Center (S$^3$TEC)", an Energy Frontier Research Center funded by the U.S. Department of Energy, Office of Science, Office of Basic Energy Science under award number DE-SC0001299.

# Notes and references

*aLaboratory of Solid State Microstructure and Innovation Center of Advanced Microstructures, Nanjing University, Nanjing 210093, China. Tel: +86 2583596595 ; E-mail: liujm@nju.edu.cn*

*bDepartment of Physics and TcSUH, University of Houston, Houston, TX 77204, USA.*

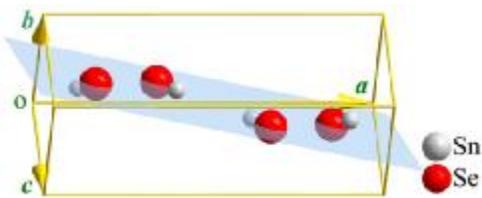

Figure1. A schematic drawing of lattice structure of low-*T* SnSe compound.

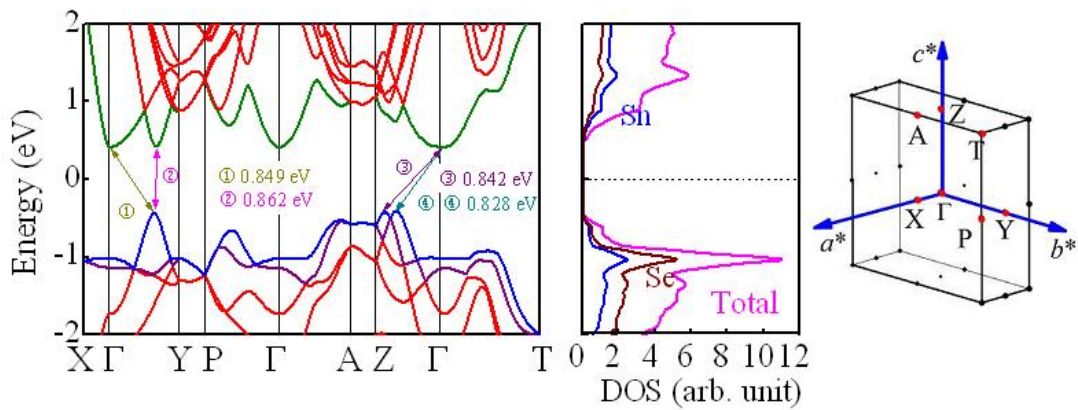

**Figure 2.** Calculated band structure and DOS spectra for Sn atoms and Se atoms as well as the total DOS for pure SnSe compound.

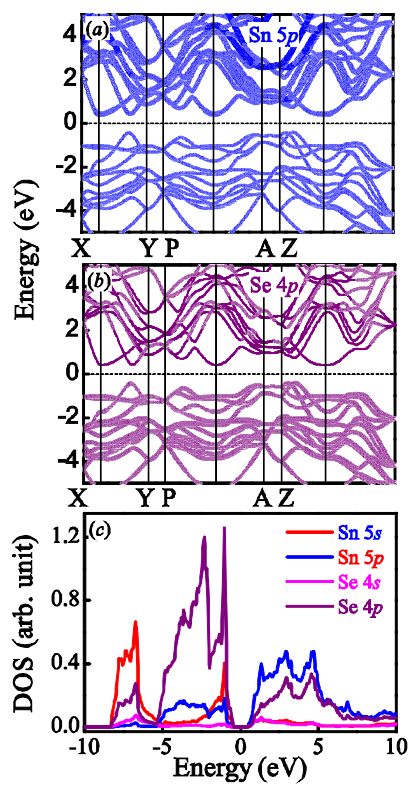

**Figure 3.** Calculated orbital-decomposed band structures for Sn 5p orbital (a) and Se 4p orbital. The projected DOS spectra for Sn 5s, Sn 5p, Se 4s, and Se 4p orbitals are plotted in (c)

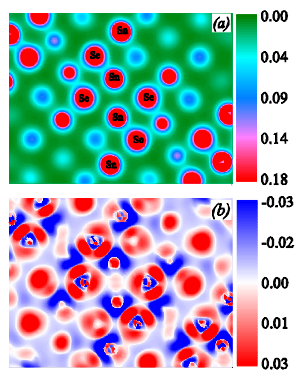

**Figure 4.** Calculated valence electron charge density (a) and electron density difference on the Se-Se-Se plane (b).

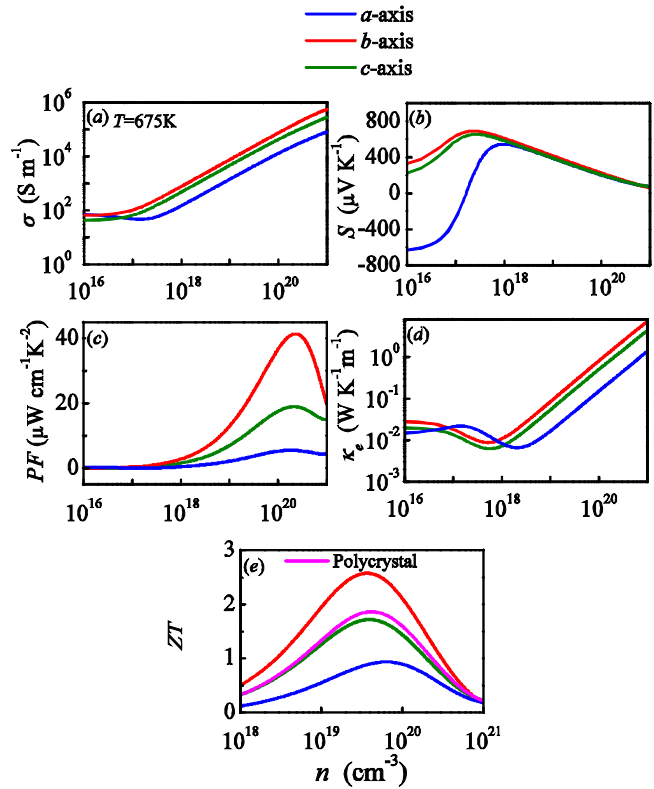

**Figure 5.** (Color online) Calculated TE parameters $S$, $\varsigma$, $k_e$, $PW$, ZT and as a function of hope-carrier density $n$ at $T$=675K.

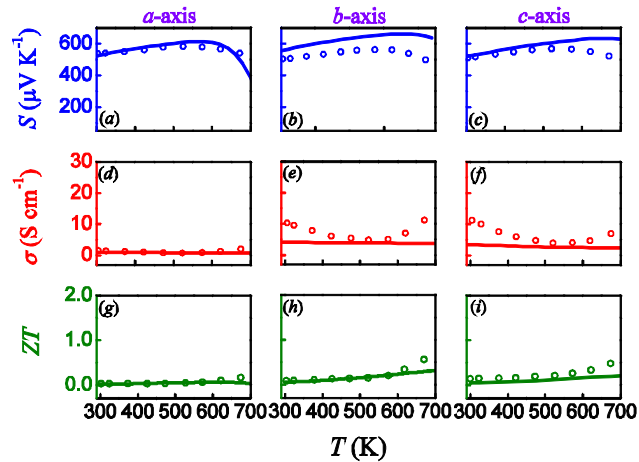

**Figure 6.** Evaluated TE parameters S (top row), $\sigma$ (middle row), and ZT (bottom row) along the three major axes, as a function of $T$, respectively. The solid lines are the calculated results and the dots are measured data extracted from Ref.[9].